\documentclass[a4paper,11pt]{article}
\pdfoutput=1 % if your are submitting a pdflatex (i.e. if you have
\usepackage{jcappub} % for details on the use of the package, please
\usepackage[T1]{fontenc} % if needed
\usepackage{bm}
\usepackage{graphicx,subfig,color}
\usepackage{amssymb,amsmath}
\usepackage{slashed}
\usepackage{array}
\usepackage{hyperref}
\usepackage{orcidlink}

\title{\boldmath Effect of dark matter interaction on hybrid star in the light of the recent astrophysical observations}

\author[a,b,1]{Suman Pal\orcidlink{0009-0000-5944-4261},\note{Corresponding author}
}
\author[a,b]{Gargi Chaudhuri \orcidlink{0000-0002-8913-0658}}
\affiliation[a]{Physics Group, Variable Energy Cyclotron Centre, 1/AF Bidhan Nagar, Kolkata 700064, India}
\affiliation[b]{Homi Bhabha National Institute, Training School Complex, Anushakti Nagar, Mumbai 400085, India}
\emailAdd{sumanvecc@gmail.com}
\emailAdd{gargi@vecc.gov.in}
\abstract{We have explored the effect of dark matter interaction on hybrid star (HS) in  the light of recent astrophysical observational constraints. The presence of dark matter is assumed to be there in both the hadron  as well as the quark sector. The dark matter particle interacts with both hadron and quark matter through the exchange of a scalar as well as a vector meson. The equation of state (EOS) of the hadron part is computed using the  NL3 version of the relativistic mean field(RMF) model, whereas the quark part is taken care of using the well-known MIT Bag model with the vector interaction. We investigate the effect of the dark matter density and the mass of the dark matter particle on various observables like mass, radius, tidal deformability of the dark matter admixed hybrid star(DMAHS).
In this study, we have  noted an intriguing aspect that is  the speed of sound in the DMAHS is insensitive to both the mass as well as the density of dark matter. We  also observe a striking similarity  in the  variation of transition mass and its corresponding radius, as well as the maximum mass of neutron stars, with dark matter density and mass. We employ observational constraints from neutron stars to narrow down the allowed range of the parameters of dark matter.}

\begin{document}
\maketitle
\flushbottom

\section{Introduction} \label{sect:intro} 
The study of neutron stars is highly interesting and challenging as it encompasses knowledge from different branches of physics, namely thermodynamics to statistical mechanics and also from general theory of relativity to particle and nuclear physics. The recent observational data on the highest as well as the lowest neutron star mass put severe constraints on the equation of state and composition of the neutron stars. Compact star equation of state is constrained by astrophysical observation such as maximum mass from PSRJ0740+6620 \cite{Fonseca:2021wxt}, NICER constraints on the radius of PSRJ0740+6620 and PSRJ0030+0451\cite{Riley:2019yda, Miller:2019cac}. The radius and mass measurement of $HESS J1731-347$\cite{2022NatAs} will be among the lightest and smallest compact objects ever detected. The gravitational wave observational\cite{LIGOScientific:2017vwq,LIGOScientific:2020zkf,LIGOScientific:2018cki} data limits the dimensionless tidal deformability of the $1.4 M_{\odot}$.

Observational data from cosmology and astrophysics have predicted the presence of a new type of undiscovered matter, called dark matter which is expected to constitute a major part of the universe.  The strongest and most straightforward evidence supporting the existence of dark matter on the scale of galaxies arises from observations of rotation curves. These curves depict the velocities at which stars and gas orbit around a galaxy's center relative to their distance from that center. Evidence also comes from the observation of the gravitational lensing and X-ray analysis of bullet clusters \cite{Randall:2008ppe,Tulin2013a}. The measurement of the cosmic microwave background anisotropy map indicates that dark matter likely makes up around 26\% of the total matter in the universe, whereas only approximately 4\% consists of baryonic matter.
The study of dark matter is one of the most interesting and intriguing topics of theoretical cosmology.
 Presently, there exist several promising particle candidates for dark matter, including weakly interacting dark matter \cite{Goldman:1989nd,Kouvaris:2007ay} and axions, among others. Exploration of the detectability of these particles have unveiled a wide array of promising experimental setups, spanning from highly precise tabletop experiments to the integration of astronomical surveys and gravitational wave observations. One must visit these \cite{Bertone:2004pz,Bramante:2023djs} for a detailed review on the recent status of knowledge about this mysterious object called dark matter.

There has been increasing attention on indications of dark matter capture within compact stellar objects\cite{Goldman:1989nd, Kouvaris:2007ay,1990PhLB238337G, Baryakhtar:2017dbj, Raj:2017wrv, Joglekar:2020liw, Bramante:2021dyx, Husain:2022brl,Deliyergiyev:2023uer} and testing dark decays of baryons in Neutron Stars \cite{McKeen:2020oyr,McKeen:2021jbh}. Neutron stars are often referred to as "graveyards" for charged dark matter because they can potentially capture and accumulate this mysterious form of matter due to their immense gravitational pull\cite{1990PhLB238337G}
The existence of dark matter inside neutron stars may be attributed to the capture or accretion of dark matter particles when the NS passes through the dark matter halo; conversion of neutrons to scalar dark matter or scalar DM production via bremsstrahlung are the other mechanisms which leads to increase in DM density inside NS. In Ref.~\cite{Brayeur} it has been pointed out that DM accumulation inside NS might be more for binary neutron star systems. The possibility of dark matter interacting with normal matter through gravitational interaction can affect the macroscopic properties of stellar objects like neutron stars. Due to their high compactness neutron star could trap the dark matter particles, which will rapidly thermalize and become accrued inside the neutron star.
Since there is a lot of uncertainty in the knowledge of the properties of dark matter(DM), different models \cite{Sen:2021wev,Kain:2021hpk,Karkevandi:2021ygv,Giangrandi:2022wht,Shakeri:2022dwg} have been considered where the nature of DM  present in neutron stars can be fermionic or bosonic. At low temperatures relevant for the cold matter in NS, bosonic (spinless dark matter particles) DM exists in the form of Bose-Einstein Condensate (BEC).  In this work, we will confine our study to fermionic dark matter only. 
Dark matter may be self-interacting
and in such cases, the masses of the DM fermion and the DM
mediators are constrained by the self-interaction constraints
from bullet cluster.

The existence of DM admixed hadronic compact stars has been thoroughly discussed in the literature \cite{Kain:2021hpk, Giangrandi:2022wht}
However relatively few works\cite{Lenzi:2022ypb}throw light on the presence of dark matter in hybrid stars, that is the neutron stars with a quark core. In the present work, we consider hybrid stars with an admixture of dark matter. We have used the very commonly used Maxwell construction\cite{Pal:2023quk} to build the hybrid star where local charge neutrality has been obeyed by the quark and the hadronic phase. For the hadronic phase, the relativistic mean field (RMF) model\cite{glendenning2012compact} have been used while for the quark phase, the MIT Bag Model\cite{chodes1974,glendenning2012compact,Pal:2023quk} modified by the vector interaction is considered. 
In the literature, the presence of dark matter inside the neutron stars is studied by considering the interaction of baryonic matter with dark matter by the exchange of  Higgs Boson interaction \cite{Panotopoulos:2017idn,Bhat:2019tnz,Das:2018frc,Dutra:2022mxl,Lenzi2022aug,Das:2021yny} as well as the exchange of scalar and vector interaction \cite{Sen:2021wev,Guha:2021njn,Sen:2022pfr}. In this work the interaction between dark matter and hadronic matter, as well as quark matter, will be considered though the exchange of one scalar and one vector meson. A distinct advantage of this model lies in the fact one can apply the single-fluid Tolman-Oppenheimer-Volkoff (TOV) equation where the DM is assumed to be uniformly distributed within NS. The other possibility 
is considering only gravitational interaction between dark matter(DM) and baryonic matter resulting in  two fluid TOV equations \cite{Tolos:2015qra}in place of the single fluid. In this scenario, one needs to specify the fraction of dark matter mass $f_D$\cite{Husain:2022brl}. 
 Another aspect is that there is a wide mass range of the dark matter particle from $10^{-12}~eV$ to $10^{15}~GeV$. In the context of the hybrid star, we are considering the values of $m_{\chi}$ from 0.1 GeV to 100 GeV.

 This paper is organized as follows. In Sec. \ref{sec:formalism} we give the
detailed formalism of the equation of state both for hadronic and quark matter with admixed dark matter. In Sec.\ref{sec:results} we show the numerical results. Finally, we summarize in Sec.~\ref{sec:conclusion}

\section{Formalism} \label{sec:formalism}

The current study focuses on exploring the properties of hybrid stars within the context of dark matter admixture.
We utilize the relativistic mean field model to describe the hadronic phase and the MIT bag model with vector interactions to describe the quark matter phase.
 
\subsection{Hadron Matter with admixed dark matter} \label{subsec:Hadron_matter}
In this work, a uniformly distributed fermionic dark matter is 
considered to be present inside a neutron star. Dark matter($\chi$) interacts via attractive scalar($\phi$) and vector dark mesons($\xi$).These meson fields also interact feebly with the nucleonic fields.  We consider the relativistic mean field (RMF) model 
 along with the dark matter\cite{Lenzi:2022ypb,Guha:2021njn,Sen:2021wev,Bhat:2019tnz,Guha:2024pnn}. As a result, the Lagrangian takes the form :
\begin{align}
\mathcal{L}_{\text{HAD}} = &\overline{\psi}(i\gamma^\mu\partial_\mu - m_N)\psi 
+ g_\sigma\sigma\overline{\psi}\psi 
- g_\omega\overline{\psi}\gamma^\mu\omega_\mu\psi - \frac{g_\rho}{2}\overline{\psi}\gamma^\mu\vec{\rho}_\mu\vec{\tau}\psi
+\frac{1}{2}(\partial^\mu \sigma \partial_\mu \sigma - m^2_\sigma\sigma^2)
 \nonumber \\
&-\frac{1}{4}\Omega^{\mu\nu}\Omega_{\mu\nu} 
+ \frac{1}{2}m^2_\omega\omega_\mu\omega^\mu 
-\frac{1}{3}bm_N(g_{\sigma}\sigma)^3-\frac{c}{4}(g_{\sigma}\sigma)^4-\frac{1}{4}\vec{\rho}^{\mu\nu}\vec{\rho}_{\mu\nu}  + \frac{1}{2}m^2_\rho\vec{\rho}_\mu\vec{\rho}^\mu,
\label{hadron_Lag}
\end{align}

\begin{align}
\mathcal{L}_{\text{total}} &= \overline{\chi}(i\gamma^\mu\partial_\mu - m_\chi)\chi
+ y_{\phi}\phi\overline{\chi}\chi +y_{\xi}\overline{\chi}\gamma^\mu\xi_\mu\chi \nonumber 
+g_{\phi}\phi\overline{\psi}\psi +g_{\xi}\overline{\psi}\gamma^\mu\xi_\mu\psi \nonumber \\
&+\frac{1}{2}m_{\xi}^2\xi_{\mu}\xi^{\mu}-\frac{1}{4}\xi_{\mu\nu}\xi^{\mu\nu}+ \frac{1}{2}(\partial^\mu \phi \partial_\mu \phi - m^2_{\phi} \phi^2) \nonumber +\mathcal{L}_{\text{HAD}}
\label{hadron_lag}
\end{align}
where $\psi$ is the nucleonic field with mass $m_N$, $\chi$ is the dark matter field with mass $m_{\chi}$; $\sigma$,~$\omega^{\mu}$ and $\vec{\rho^{\mu}}$ are the scalar, vector and isovector meson
fields, respectively. 
Here $\mathcal{L}_{HAD}$ is the pure hadronic part. $\mathcal{L}_{HAD}$ contains five coupling constants :~$g_{\sigma},~g_{\omega},~g_{\rho}$,~b and c. The coupling constants are determined by fixing the nuclear saturation properties. In this work, we are using NL3 parametrization with the following  nuclear saturation parameters: nuclear saturation density ($\rho_0)=0.148 fm^{-3}$, binding energy per nucleon
($\frac{BE}{A})=-16.24 ~\text{MeV}$, the incompressibility coefficient ($K_{sat})=271.5 ~\text{MeV}$,~effective mass ($\frac{m^*}{m})=0.55$,~ symmetry energy at saturation ($E_{sym})=37.29~\text{MeV}$~ and the slope of the symmetry energy at saturation ($L_{sym})=118.2~\text{MeV}$ based on Ref \cite{Lalazissis:1996rd,Fattoyev:2010mx,Chen:2014sca}. The values of the hadronic parameters are given in the table \ref{tab:1}. The parameters of the dark sector are given in the sub-section \ref{sub:dark_params}.
\begin{table}[!ht]
\caption{The coupling constants of the NL3 model for the hadronic sector.}
\setlength{\tabcolsep}{10pt}
\begin{tabular}{cccccccc}
\hline
\hline
$m_{\sigma}$ (MeV) &$g_{\sigma}^2$ &$m_{\omega}$(MeV)& $g_{\omega}^2$ & $m_{\rho}$ (MeV) & $g_{\rho}^2$ & $b$ & c  \\
\hline
508.194 & 104.3871  & 782.5  & 165.5854 & 763 & 79.6 & 0.00205& -0.0026508 \\
\hline
\end{tabular}
\label{tab:1}
\end{table}

We use the relativistic mean field approximation, replacing the meson field with their mean fields. Our lagrangian includes a total of five meson fields, namely, $\sigma$, $\omega$, $\rho$, $\phi$ and $\xi$. The meson field equations are :
\begin{equation}
   \begin{aligned}
    &m_{\sigma}^2\sigma_0=g_{\sigma}\sum_{f=p,n}\rho_{sf}-bm(g_{\sigma}\sigma_0)^2-cg_{\sigma}(g_{\sigma}\sigma_0)^3 ~,\\
    &m_{\omega}^2\omega_0=g_{\omega}\sum_{f=p,n}\rho_f~, \\
    &m_{\rho}^2\rho_{30}=\frac{1}{2}\sum_{f=p,n}\tau_3\rho_f~, \\
    &m_{\phi}^2\phi_0=g_{\phi}\sum_{f=p,n}\rho_{sf}+y_{\phi}\rho_{s\chi} ~,\\
    &m_{\xi}^2\xi_0=g_{\xi}\sum_{f=p,n}\rho_f+y_{\xi}\rho_{\chi}~.
   \end{aligned}
\end{equation} 
Due to the interaction with scalar mesons  the mass of the nucleons and dark matter particles are modified as 
\begin{equation}
    \begin{aligned}
        &m_N^*=m_N-g_{\sigma}\sigma_0-g_{\phi}\phi_0 ~,\\
        & m_{\chi}^*=m_{\chi}-y_{\phi}\phi_0~.\\
    \end{aligned}
\end{equation} 
The scalar and vector densities are expressed as follows: 
\begin{equation}
\begin{aligned}
    &\rho_{sf}=\frac{m_N^*}{\pi^2}\int_0^{k_f} \frac{k^2}{\sqrt{(k^2+(m_N*)^2)}}   dk ~,\\
    &\rho_{s\chi}=\frac{m_{\chi}^*}{\pi^2}\int_0^{k_{f_{\chi}}} \frac{k^2}{\sqrt{(k^2+(m_{\chi}^*)^2)}}   dk ~,\\
    &\rho_f=\frac{k_f^3}{3\pi^2}~,\\
    &\rho_{\chi}=\frac{k_{f_{\chi}}^3}{3\pi^2}~.
\end{aligned}
\end{equation} 
Here $\rho_f$ is the baryon density for the proton and neutron and $\rho_{\chi}$ is dark matter density corresponding to the fermi momentum $k_{f_{\chi}}$.
Due to the interaction of vector mesons, the chemical potential of the nucleon and dark matter particle gets modified as :
\begin{equation}
\begin{aligned}
& \mu_f=\sqrt{k_f^2+m_N^*}+g_{\omega}\omega_0+\tau_{3}\frac{1}{2}g_{\rho}\rho_{30}+g_{\xi}\xi_0 ~,\\
&    \mu_{\chi}=\sqrt{k_{_f{\chi}}^2+(m^*_{\chi})^2}+g_{\xi}\xi_0~. \\
\end{aligned}
\end{equation} 
The energy and pressure are given by the following expressions as
\begin{equation} 
\begin{aligned}
        \varepsilon=&\frac{1}{2}m_{\sigma}^2\sigma_0^2+\frac{1}{2}m_{\omega}^2\omega_0^2+\frac{1}{2}m_{\rho}^2\rho_{30}^2+\frac{1}{2}m_{\phi}^2\phi_0^2+\frac{1}{2}m_{\xi}^2\xi_0^2+\sum_{f=p,n} \int_0^{k_f}\frac{\gamma_f}{2\pi^2}\sqrt{k_f^2+(m_N^*)^2}k^2 dk+\\&\int_0^{k_{f_{\chi}}}\frac{\gamma_f}{2\pi^2}\sqrt{k_f^2+(m_{\chi}^*)^2}k^2 dk+\frac{1}{3}bm_N(g_{\sigma}\sigma)^3+\frac{c}{4}(g_{\sigma}\sigma)^4~,\\
P=&-\frac{1}{2}m_{\sigma}^2\sigma_0^2+\frac{1}{2}m_{\omega}^2\omega_0^2+\frac{1}{2}m_{\rho}^2\rho_{30}^2-\frac{1}{2}m_{\phi}^2\phi_0^2+\frac{1}{2}m_{\xi}^2\xi_0^2+\sum_{f=p,n} \int_0^{k_f}\frac{\gamma_f}{6\pi^2}\frac{k^4}{\sqrt{k_f^2+(m_N^*)^2}}dk+\\&\int_0^{k_{f_{\chi}}}\frac{\gamma_f}{6\pi^2}\frac{k^4}{\sqrt{k^2+(m_{\chi}^*)^2}} dk-\frac{1}{3}bm_N(g_{\sigma}\sigma)^3-\frac{c}{4}(g_{\sigma}\sigma)^4~.\\
\end{aligned}
\end{equation}
%========================================
\begin{figure*}[htp]
    \centering
        \includegraphics[width=0.45\textwidth]{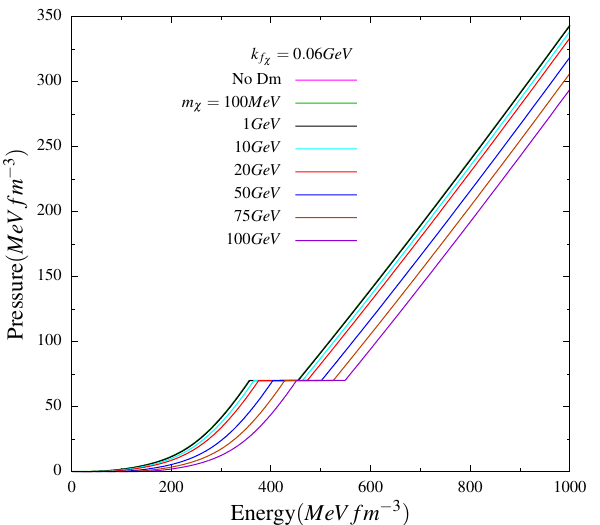}
    \includegraphics[width=0.45\textwidth]{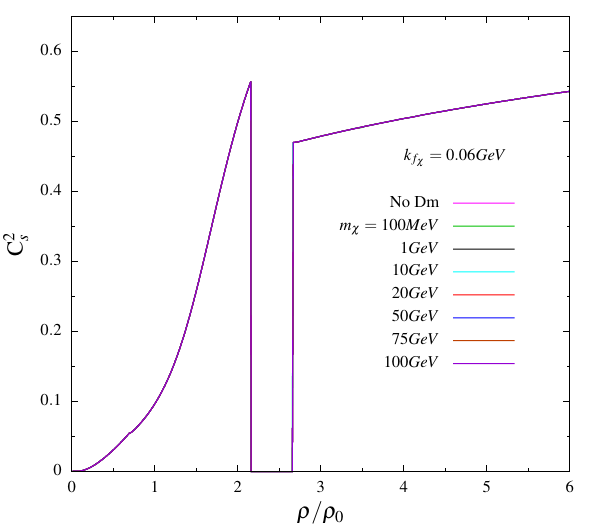} 
    \includegraphics[width=0.45\textwidth]{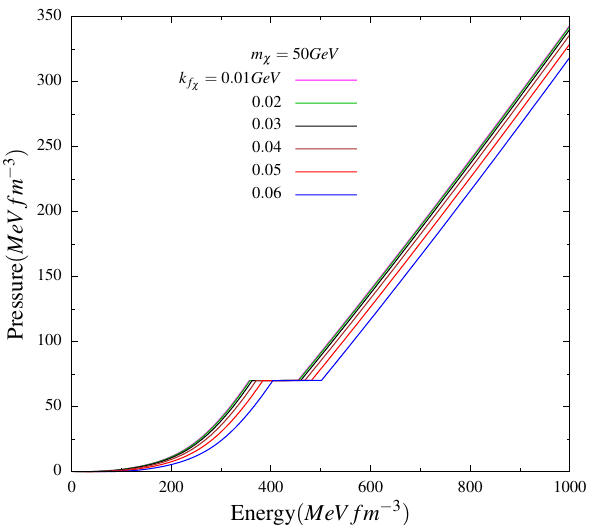}
    \includegraphics[width=0.45\textwidth]{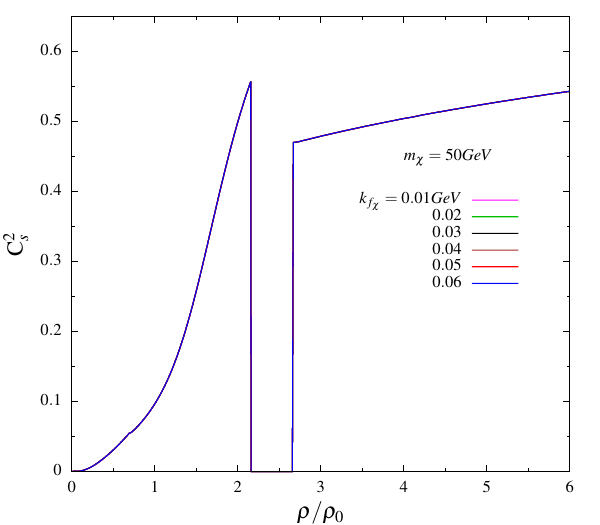}
\caption{Pressure vs energy density (left panel) and speed of sound vs baryon density (right panel) for a dark matter admixed  hybrid star;  different values of $m_{\chi}$ with fixed $k_{f_{\chi}}=0.06$ GeV(upper panel) and  different values of $k_{f_{\chi}}$ with fixed $m_{\chi}=50$ GeV (lower panel)}
\label{fig:eos_cs2_mchi}  
\end{figure*} 
%============================================

\subsection{Quark Matter admixed dark matter}\label{sec:Quark_matter}
We utilize the MIT bag model with vector interaction to characterize the combination of quark matter and dark matter. 
We explore feeble interaction between fermionic dark matter and strange quark matter within a quark star, mediated by scalar and vector dark bosons. As a result, the Lagrangian takes the form \cite{Guha:2021njn,Sen:2021wev,Sen:2022pfr}:
\begin{equation}
\begin{aligned}
\mathcal{L} = &\sum_{f=u,d,s} \left[ \bar{\psi}_f (i\gamma_{\mu}\partial^{\mu} - m_f) \psi_f - B \right] \Theta(\bar{\psi_f}\psi_f)- \sum_{f=u,d,s} \left[ \bar{\psi}_f \gamma_{\mu} (g_{V}V^{\mu} - g_{\xi}\xi^{\mu}) \psi_f \right] \Theta(\bar{\psi_f}\psi_f) \\
&+ \frac{1}{2}m_V^2 V_{\mu} V^{\mu} - \frac{1}{4}V_{\mu\nu} V^{\mu\nu} + \bar{\psi}_e (i\gamma_{\mu}\partial^{\mu} - m_e) \psi_e + \bar{\chi} \left( i\gamma_{\mu}\partial^{\mu} - y_{\xi}\gamma_{\mu}\xi^{\mu} - (m_{\chi} - y_{\phi}\phi) \right) \chi \\
&+ \frac{1}{2}m_{\xi}^2 \xi_{\mu} \xi^{\mu} - \frac{1}{4}\xi_{\mu\nu} \xi^{\mu\nu}+\frac{1}{2}(\partial^\mu \phi \partial_\mu \phi - m^2_{\phi} \phi^2) \\
\end{aligned}
\end{equation}

Here quark interaction is mediated by one attractive scalar dark meson ($\phi$) and two vectors (V (normal)and $\xi$(dark)) repulsive mesons. The bag constant (B) is added to the lagrangian to reflect the quark confinement. The normal vector interaction coupling constant($g_V$)  is scaled with $G_V=\left(\frac{g_V}{m_V}\right)^2$, where $m_V$ is the mass of the vector meson (V). The interaction between the quark sector and dark boson is taken to be feeble as mentioned in \cite{Sen:2022pfr}. In this work, we consider  the masses of the quarks to be $m_u=2.16$ \text{MeV}, $m_d=4.67$ \text{MeV} and $m_s=93.4$ \text{MeV} and vector coupling constant $G_V=0.35~fm^{-2}$.
After the mean-field approximation, the equation of motion of the meson fields are given as 
\begin{equation}
    \begin{aligned}
    m_V^2 V_0=&g_V\sum_{f=u,d,s}\rho_f~,\\
   m_{\xi}^2\xi_0= &g_{\xi}\sum_{f=u,d,s}\rho_f+y_{\xi}\rho_{\chi}~,\\
   m_{\phi}^2\phi_0=&g_{\phi}\sum_{f=u,d,s}\rho_{sf}+y_{\phi}\rho_{s\chi}~.\\
       \end{aligned}
\end{equation}  
The expressions for the scalar and vector densities for the quark admixed dark matter configuration are as follows:
\begin{equation}
\begin{aligned} 
    &\rho_{sf}=\frac{m_f^*}{\pi^2}\int_0^{k_f} \frac{k^2}{\sqrt{(k^2+(m_f*)^2)}}  dk ~, \\
    &\rho_{s\chi}=\frac{m_{\chi}^*}{\pi^2}\int_0^{k_{f_{\chi}}} \frac{k^2}{\sqrt{(k^2+(m_{\chi}^*)^2)}}  dk ~ , \\ 
    &\rho_f=\frac{k_f^3}{\pi^2}~,\\
    &\rho_{\chi}=\frac{k_{f_{\chi}}^3}{3\pi^2}~.
\end{aligned}
\end{equation} 

In the MIT bag model, the quark mass is conventionally treated as constant. However, when incorporating the scalar boson ($\phi$), both the quark mass and the mass of dark matter undergo modifications.
\begin{equation}
    \begin{aligned}
           m_f^*=&m_f-g_{\phi}\phi_0~,\\
    m_{\chi}^*=&m_{\chi}-y_{\phi}\phi_0~.\\ 
    \end{aligned}
\end{equation}
Because of the vector interaction, there are modifications to the chemical potentials of both quarks and dark matter.
\begin{equation}
    \begin{aligned}
        &\mu_f=\sqrt{k_f^2+(m_f^*)^2}+g_VV_0+g_{\xi}\xi_0~,\\
        &\mu_{\chi}=\sqrt{k_{_f{\chi}}^2+(m_{\chi}^*)^2}+y_{\xi}\xi_0~.
    \end{aligned}
\end{equation}
The energy and pressure expressions are given by
\begin{equation}
\begin{aligned}
\varepsilon &= \frac{1}{2}m_{V}^2V_0^2 + \frac{1}{2}m_{\xi}^2\xi_0^2 + \frac{1}{2}m_{\phi}^2\phi_0^2+B + \frac{\gamma_f}{2\pi^2}\sum_{f=u,d,s} \int_0^{k_f}\sqrt{k^2+(m_f^*)^2}k^2 dk \\
&+ \frac{\gamma_e}{2\pi^2}\int_0^{k_e}\sqrt{k^2+m_{e}^2}k^2 dk + \frac{\gamma_{\chi}}{2\pi^2}\int_0^{k_{f_{\chi}}}\sqrt{k^2+(m_{\chi}^*)^2}k^2 dk ~,\\
P &= \frac{1}{2}m_{V}^2V_0^2 + \frac{1}{2}m_{\xi}^2\xi_0^2 - \frac{1}{2}m_{\phi}^2\phi_0^2-B + \frac{\gamma_f}{6\pi^2}\sum_{f=u,d,s} \int_0^{k_f}\frac{k^4}{\sqrt{k^2+(m_f^*)^2}} dk \\
&+ \frac{\gamma_e}{6\pi^2}\int_0^{k_e}\frac{k^4}{\sqrt{k^2+m_{e}^2}} dk + \frac{\gamma_{\chi}}{6\pi^2}\int_0^{k_{f_{\chi}}}\frac{k^4}{\sqrt{k^2+(m_{\chi}^*)^2}} dk~.
\end{aligned}
\end{equation}

\begin{figure*}[htp]
    \centering
        \includegraphics[width=0.45\textwidth]{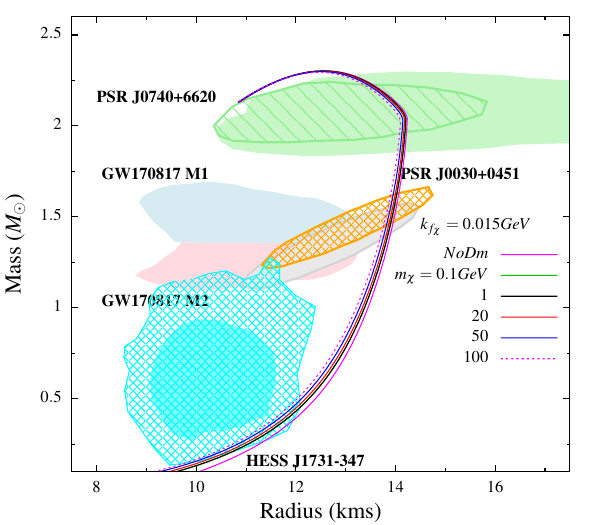}
    \includegraphics[width=0.45\textwidth]{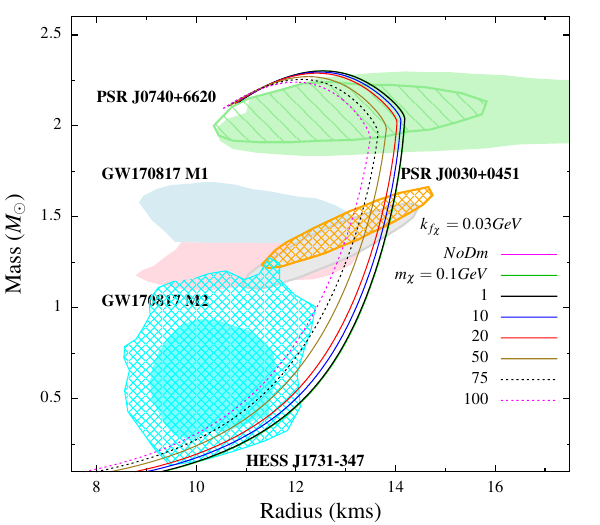}
        \includegraphics[width=0.45\textwidth]{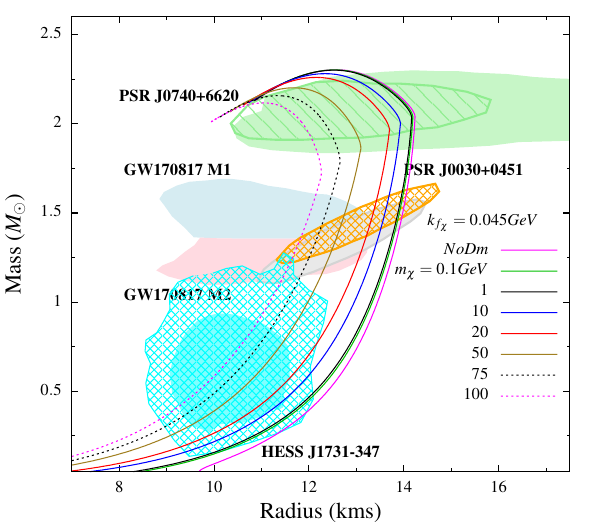}
    \includegraphics[width=0.45\textwidth]{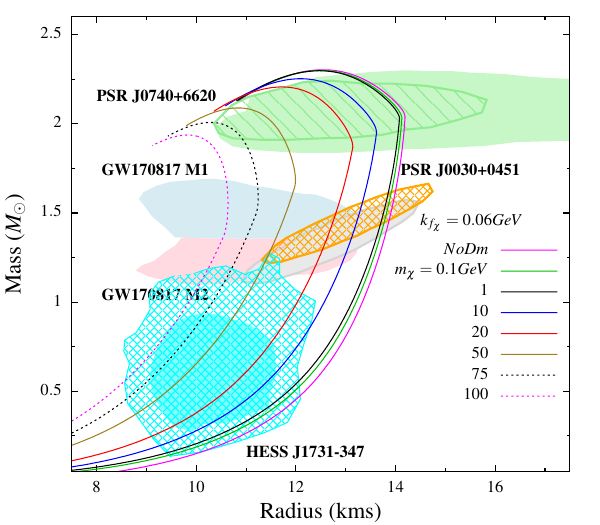}

\caption{Mass-radius diagram corresponding to the different values of $m_{\chi}$ with fixed $k_{f_{\chi}}=0.015, 0.03, 0.045$ and $0.06$ GeV. The restrictions on the M-R plane from GW170817, the NICER experiment for PSR J0030+0451 PSR J0740+6620, and HESSJ1731-347 have been incorporated. }
\label{fig:MR_HS}  
\end{figure*}  

%==============================================================
\subsection{Parameter configuration for the dark sector}\label{sub:dark_params} In this work, we assume that the average dark matter number density inside the neutron star is $10^3$ times smaller than the saturated nuclear matter density, and we assume that the Fermi momentum of the dark matter remains constant throughout the neutron star. % $\frac{M_{\chi}}{M_{NS}}=\frac{1}{6}$.$\rho_{DM} \sim 10^{-3}\rho_0$,$k_{DM} \sim 0.033 GeV$. 
In our calculations, we explore a range of dark matter Fermi momenta, with values ranging from approximately $0.01$ to $0.06$ GeV.
 
We consider the dark matter self-interaction constraints  from the bullet cluster, for the DM fermions of mass $\frac{\sigma_T}{m_{\chi}} = (0.1 - 10) \, \text{cm}^2 \, \text{g}^{-1}$\cite{Tulin2013a} 
Assuming the cross-section of the self-interaction to be
\begin{equation} \label{cross_section_1}
    \sigma_{\chi} \sim 5 \times 10^{-23} \left(\frac{\alpha_{x}}{0.01}\right)^2 \left(\frac{m_{\chi}}{10 \, \text{GeV}}\right)^2 \left(\frac{10 \, \text{MeV}}{m_{x}}\right)^2 \, \text{cm}^2
\end{equation}
where 
$\alpha_x = \frac{g_{x}^2}{4\pi} \quad \text{and} \quad G_{x} = \left(\frac{y_{x}}{m_{x}}\right)^2$
where x may be the scalar and vector mediator. We have taken $m_x=10$ MeV
, then 
\begin{equation}
    \frac{\sigma_{\chi}}{m_{\chi}} = 1.2 \times 10^{-4} \left(\frac{G_{x}}{1 \, \text{fm}^2}\right) \left(\frac{m_{\chi}}{1 \, \text{GeV}}\right) \, \text{cm}^2/\text{gm}
\end{equation}

For the lower bound, we take $\frac{\sigma_{\chi}}{m_{\chi}} = 0.1 \, \text{cm}^2\text{gm}^{-1}$, which gives rise to $y_{x} = 0.2687\left(\frac{1 \text{GeV}}{m_{\chi}}\right)^{\frac{1}{4}}$.

For the upper bound, $\frac{\sigma_{\chi}}{m_{\chi}} = 10 \, \text{cm}^2\text{gm}^{-1}$ gives rise to $y_{x} = 1.511\left(\frac{1 \text{GeV}}{m_{\chi}}\right)^{\frac{1}{4}}$.

The final limit on the dark matter to the dark scalar and vector boson couplings reads as
\begin{equation}
 0.2687\left(\frac{1 \text{GeV}}{m_{\chi}}\right)^{\frac{1}{4}} \leq y_{x} \leq 1.511\left(\frac{1 \text{GeV}}{m_{\chi}}\right)^{\frac{1}{4}}
\end{equation}
In this study, we have adopted the values of $g_{\xi}$ and $g_{\phi}$ to be approximately $10^{-4}$, as previously mentioned in reference \cite{Sen:2021wev,Sen:2022pfr}. The values of $m_{\chi}$ and $y_x$ are given in  table~\ref{tab:2}. For the nuclear sector, we  have employed the NL3 parametrization, while for the quark sector, we  have employed the vector bag model with a bag constant of $B^{1/4}=160$ MeV. 
{\color{red} 
\begin{table}[!ht] 
\centering
\caption{The coupling constants of the dark sector.}
\setlength{\tabcolsep}{10pt}
\begin{tabular}{cc}
\hline
\hline
$m_{\chi}$ (GeV) &$y_{x}$   \\
\hline
0.1& 2.2\\
1.0& 1.0 \\
10 & 0.8 \\
20& 0.7 \\
50& 0.4  \\
100& 0.1 \\
\hline
\end{tabular}
\label{tab:2}
\end{table}

}

\section{Results} \label{sec:results}
\begin{figure*}[htp]
    \centering
        \includegraphics[width=0.32\textwidth]{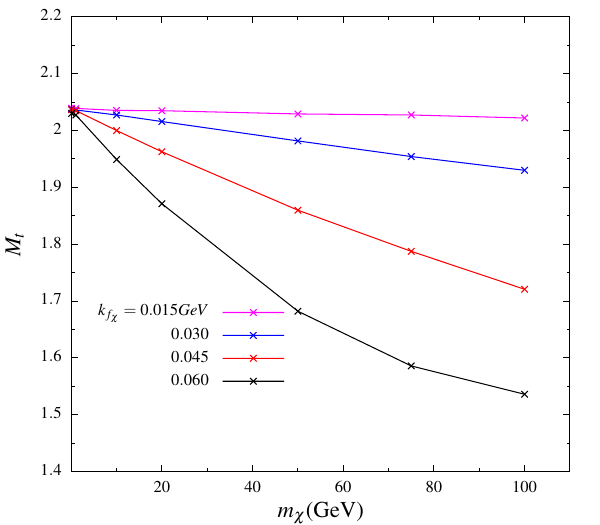}
    \includegraphics[width=0.32\textwidth]{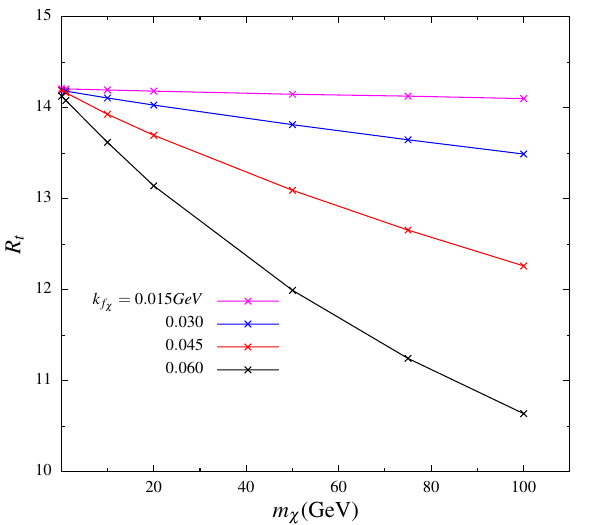}
     \includegraphics[width=0.32\textwidth]{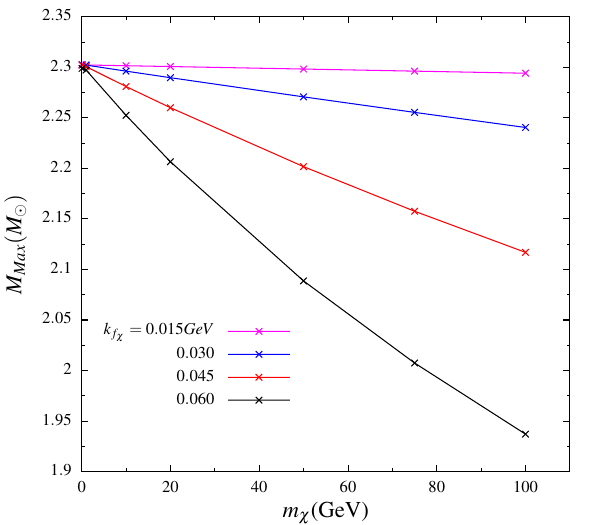} 
\caption{ The variation of $M_{t}$(left), $R_t$(middle) and $M_{Max}$(right) with $k_{f_{\chi}}$ for the different values of $m_{\chi}$.}
\label{fig:fit_reln_MR_M_chi}  
\end{figure*}   

%===============  Fig 6============================
\begin{figure*}[htp]
    \centering
        \includegraphics[width=0.32\textwidth]{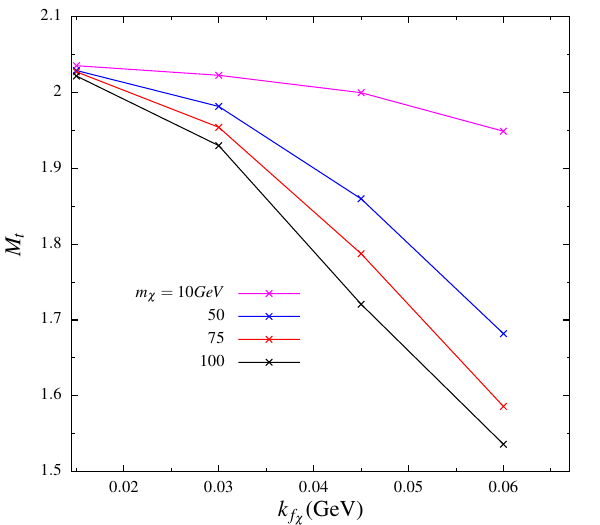}
    \includegraphics[width=0.32\textwidth]{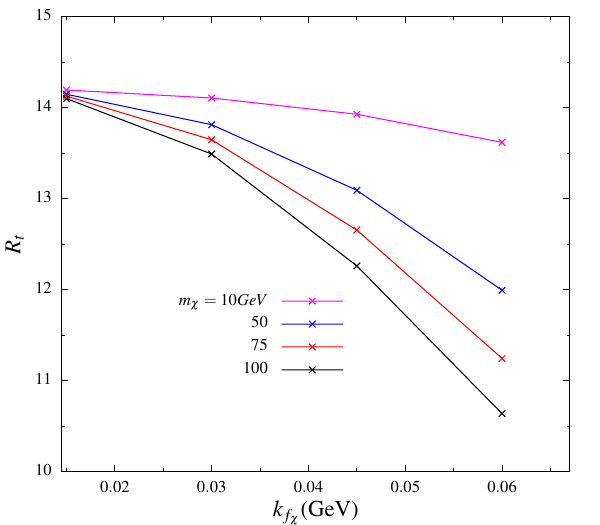}
     \includegraphics[width=0.32\textwidth]{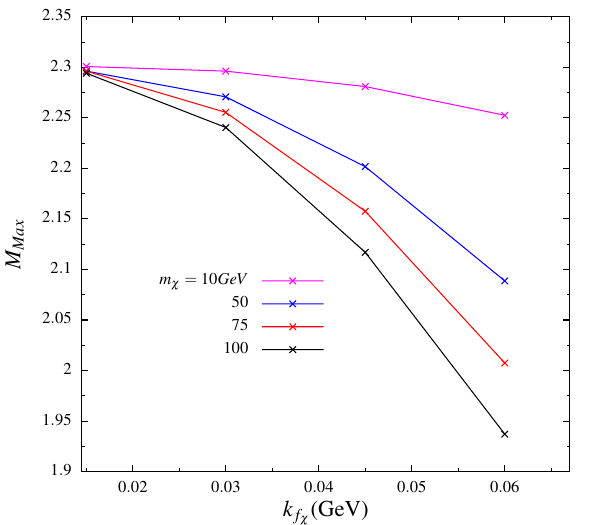} 
\caption{  The variation of $M_{t}$(left), $R_t$(middle) and $M_{Max}$(right) with $m_{\chi}$ for the different values of $k_{f_{\chi}}$.}
\label{fig:fit_reln_MR_k_chi}  
\end{figure*}  
In this work, the phase transition from hadronic to quark phase is obtained with Maxwell construction where
the pressure and baryon chemical potential of the individual charge-neutral phases are equal. 
In Fig.~\ref{fig:eos_cs2_mchi} ( upper left panel) we plot the equation of state (EOS) for different values of the mass of the DM particle( $m_{\chi}$) for a fixed value of the dark matter Fermi momentum $k_{f_{\chi}}$ (=0.06 GeV).  The results are pretty close to each other for the range of the masses used. In the upper right panel, we have plotted the square of the velocity of sound $C_s^2$ for dark matter and it has been observed that this observable is insensitive to $m_{\chi}$ and hence $C_s^2$ remains unchanged irrespective of the values chosen. In the lower panel, we plot the EOS for different 
$k_{f_{\chi}}$ keeping the mass of DM particle $m_{\chi}$ fixed.  Here also the plots are pretty close to each other for different values. In the right lower panel, we plot the corresponding $C_s^2$. It is once again observed that this is 
insensitive to DM parameters and cannot discriminate among the different values. From this figure, it seems that the effect of the dark matter is negligible, but this is not true. The reason being that the contribution from the dark matter kinetic energy term is much greater than the kinetic pressure integral. The kinetic energy and pressure integral do not vary much with baryon density. Therefore change in the pressure energy slope is negligible and it does not affect the speed of sound.

\begin{figure*}[htp]
\centering
\includegraphics[width=0.45\textwidth]{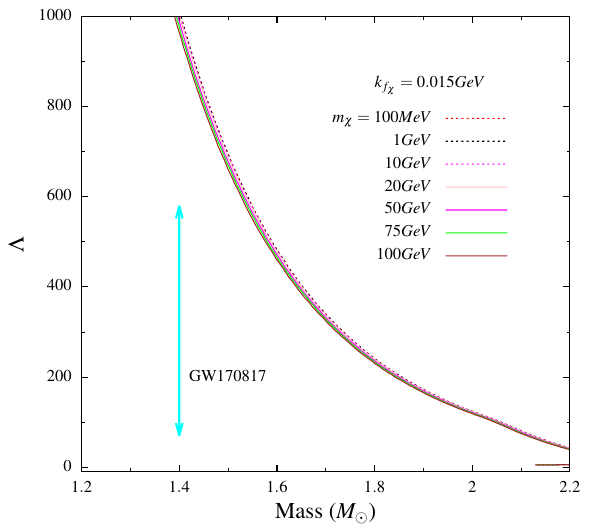}
\includegraphics[width=0.45\textwidth]{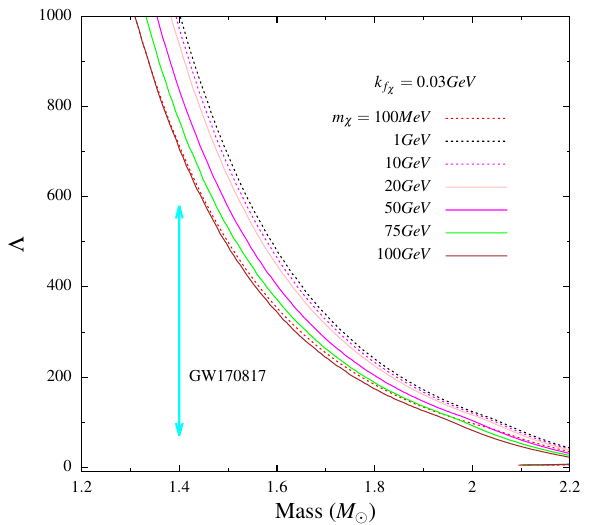}
\includegraphics[width=0.45\textwidth]{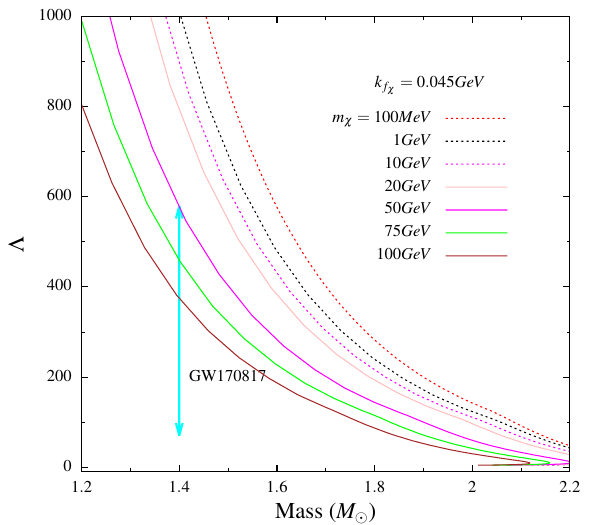}
\includegraphics[width=0.45\textwidth]{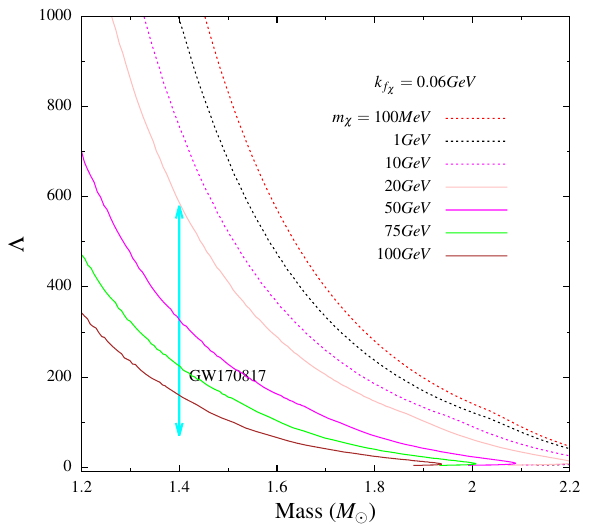}
\caption{Tidal deformability-Mass  diagram  with different values of $m_{\chi}$ with fixed $k_{f_{\chi}}=$0.015(upper left), 0.03(upper right), 0.045(lower left) and 0.06 (lower right)  GeV. The restrictions on the $M-\Lambda$ plane from GW170817 have been incorporated. }
\label{fig:lambda_HS}  
\end{figure*}  
In Fig.~\ref{fig:MR_HS},  we plot the mass-radius of the hybrid stars for scaled coupling ($G_V$= 0.35 $fm^{-2}$ )for different values of the dark matter Fermi momentum $k_{f_{\chi}}$. For fixed $k_{f_{\chi}}$, in each figure we have shown the effect of the mass of the dark matter particle $m_{\chi}$ on the mass and radius of the HS. In each figure, we have also indicated the plot for no dark matter case in order to clearly demonstrate the difference. When the dark matter fermi momentum ($k_{f_{\chi}}$) (upper left figure) is low, there is not much difference between the plots as we vary the mass of the dark matter particle. This case of $k_{f_{\chi}}=0.015~\text{GeV}$ satisfies the observational constraints from  PSR J0030+0451 and PSR J0740+6620 but fails to satisfy those from Gravitational Wave observation and HESS J1731-347. 
As we increase the value of $k_{f_{\chi}}$, the effect of the mass of the dark matter particle $m_{\chi}$  becomes more and more evident as seen in the other plots of Fig.~\ref{fig:MR_HS}. With the increase in the mass of the dark matter particle, the transition mass ($M_t$) (where hadron to quark phase transition takes place), the maximum mass ($M_{max}$)  as well as the radius $R_t$ (radius corresponding to $M_t$) of the HS decreases. The results for higher mass DM particles satisfy the constraints from GW170817  as well as that from HESS J1731-347.  Thus the understanding of  the nature of the lightest observed neutron star $HESSJ1731-347$, can be addressed 
  from the possible scenario of the dark matter admixed hybrid star configuration. From Fig.~\ref{fig:MR_HS}. we see that massive dark matter and high Fermi momentum dark matter particles are more favorable to satisfy the $HESSJ1731-347$ constraints. 
It is also noticed that $M_t$, $M_{max}$, as well as $R_t$ of the HS decreases with $k_{f_{\chi}}$. These features will be further demonstrated in the next figures.
In Fig.~\ref{fig:fit_reln_MR_M_chi}, we show the variation of  $M_t$, $R_t$  and $M_{max}$ with $m_{\chi}$ for different values of $k_{f_{\chi}}$. When  $k_{f_{\chi}}$ is very small (0.015 GeV), there is almost negligible or no change of these parameters with mass of the DM particle. The decrease of these parameters becomes more noticeable as we increase the value of $k_{f_{\chi}}$ and the change is most for the highest value (0.06 GeV) of  $k_{f_{\chi}}$. For a fixed value of $k_{f_{\chi}}$, it is seen that the 
 nature of  variation of all the three parameters,  $M_t$, $M_{max}$, as well as $R_t$, is very similar as seen from Fig.~\ref{fig:fit_reln_MR_M_chi}. The fitted functional relationship of each of $M_t, R_t~\text{and}~M_{Max}$ with $m_{\chi}$ is quite similar for a fixed value of dark matter fermi momentum. We demonstrate here only one such fitted functional form for the value of $k_{f_{\chi}}=0.06~\text{GeV}$ as given below
\begin{equation}
    \begin{aligned}
        M_t=&2.3772 \times 10^{-5}(m_{\chi}^2+m_{\chi})+ 2.03457~exp\left(-\frac{m_{\chi}}{221.599}\right) \\
        R_t=&8.1067 \times 10^{-5}(m_{\chi}^2+m_{\chi})+ 14.1205~exp\left(-\frac{m_{\chi}}{276.16}\right) \\
        M_{Max}=&7.4639 \times 10^{-6}(m_{\chi}^2+m_{\chi})+ 2.3004~exp\left(-\frac{m_{\chi}}{473.882}\right) \\
    \end{aligned}
\end{equation} 
%where A, B and C are different constants and coefficients of this   parametrization as seen from the equations above.  
In Fig.~\ref{fig:fit_reln_MR_k_chi} we plot $M_t$, $R_t$  and $M_{max}$ as a function of $k_{f_{\chi}}$ for fixed values of the DM particle mass $m_{\chi}$. It is observed that all the three parameters follow same pattern of variation with $k_{f_{\chi}}$ irrespective of the dark matter particle mass $m_{\chi}$.
The fitted relationship of $M_t$, $R_t$, $M_{Max}$ with $k_{f_{\chi}}$ are given below  for $m_{\chi}=100 GeV$
\begin{equation}
    \begin{aligned}
        M_t=& 6281.98~k_{f_{\chi}}^3-809.756~k_{f_{\chi}}^2+19.9982~k_{f_{\chi}}+1.88916 \\
        R_t=& 10172.8~k_{f_{\chi}}^3-2246.67~k_{f_{\chi}}^2+42.3444~k_{f_{\chi}}+13.985 \\ 
        M_{Max}=& 567.407~k_{f_{\chi}}^3-203.533~k_{f_{\chi}}^2+4.618~k_{f_{\chi}}+2.26992 \\
    \end{aligned}
\end{equation} 
\begin{figure*}[htp]
    \centering 
     \includegraphics[width=0.45\textwidth]{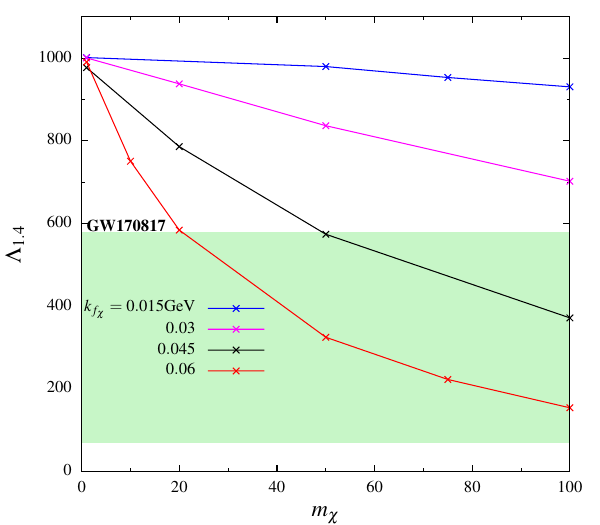}
    \includegraphics[width=0.45\textwidth]{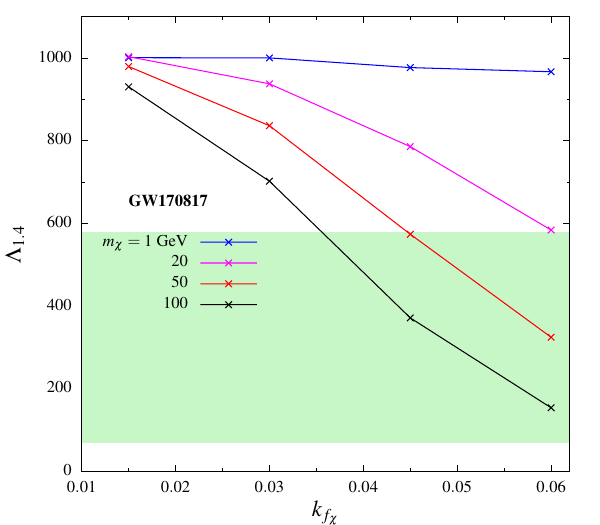}
\caption{ The variation of $\Lambda_{1.4}$ with  $m_{\chi}$ for different values of $k_{f_{\chi}}$(left panel) and $k_{\chi}$ for different values of $m_{\chi}$(right panel).}
\label{fig:fit_reln_lambda1.4}  
\end{figure*}

It is observed that all these variations follow the same parametrization with the coefficients of the expansion being different for different values of $m_{\chi}$. This is indeed a remarkable observation of the effect of the dark matter interaction on HS properties.

In  Fig.~\ref{fig:lambda_HS}, we show the variation of tidal deformability with mass. In each subfigure, we demonstrate the variation of tidal deformability with the mass of the dark matter particle keeping the fermi momentum $k_{f_{\chi}}$ fixed. For low $k_{f_{\chi}}$ (upper left figure), it is observed that tidal deformability is not sensitive to the mass of the dark matter particle $m_{\chi}$ and also the constraints from GW170817 is not satisfied. As one increases the value of the Fermi momentum (upper right figure), $\Lambda$ could better discriminate between the masses of the dark matter particle and those with higher masses do satisfy the constraints from the gravitational wave (GW) observation. From the lower panel, a similar trend continues to be observed with further increase in the value of $k_{f_{\chi}}$. The sensitivity of the tidal deformability parameter on the mass of the dark matter particle increases reaching its maximum for the maximum value of $k_{f_{\chi}}$ used which is equal to 0.06 GeV. The range of $m_{\chi}$ values satisfying the constraint from GW170817 increases with $k_{f_{\chi}}$.  Masses of dark matter particle $m_{\chi}$ spanning the range of 20 to 100 GeV obey the constraint for  $k_{f_{\chi}}=0.06$ GeV. A similar trend has also been observed from the mass-radius plots in Fig.~\ref{fig:MR_HS}. 
We next study the effect $m_{\chi}$ and $k_{f_{\chi}}$ on $\Lambda_{1.4}$ as shown in the Fig.~\ref{fig:fit_reln_lambda1.4}. The value of  $\Lambda_{1.4}$ decreases with $m_{\chi}$ and $k_{\chi}$. From the left panel of Fig.~\ref{fig:fit_reln_lambda1.4},  $\Lambda_{1.4}$ calculations reveal that  the DM fermi momenta values $k_{f_{\chi}}=0.015$ GeV and $k_{f_{\chi}}=0.03$ GeV are restricted by GW170817 observation. The lower values of $m_{\chi}$ are also restricted. In the right panel of the Fig.~\ref{fig:fit_reln_lambda1.4} it is inferred that  DM particle mass  up to $m_{\chi}=20$ GeV and fermi momenta up to $k_{f_{\chi}}=0.03$~GeV are not allowed by the $\Lambda_{1.4}$ observation.
We also  observe that from the GW170817 tidal deformability constraint, the lower values of $k_{f_{\chi}}$ and $m_{\chi}$ are ruled out.

%============= Fig 7=======================

\section{Summary and conclusion }
\label{sec:conclusion} 
In this work, we have investigated the dark matter accreted hybrid stars resulting in  the formation of  DM admixed hybrid stars.  We use the RMF model with$NL3$ parametrization for the hadronic sector and the MIT bag model with vector interactions for the quark sector; phase transition being achieved by using Maxwell construction. We have studied the effect of the dark matter parameters $m_{\chi}$ and $k_{f_{\chi}}$ on the equation of state, speed of sound, mass-radius relations, and tidal deformability of the HS. The important finding of our result is that the speed of sound is insensitive to the dark matter parameters ($m_{\chi}$~and~$k_{f_{\chi}}$). In the $M-R$ diagram, we find that PSRJ0740+6620 data is satisfied by most of our chosen parameters of DM except for the more massive and large momentum cases. PSRJ0030+0451  is satisfied by the all parameter set of our chosen model. We have constrained the mass and momentum range of the DM particle from the GW170817 constraint. We have also explored to find the possible nature of HESSJ1731-347 in the light of our dark matter admixed hybrid star model. In order to understand the effect of the dark matter parameters, we plot $M_t$,$R_t$, and $M_{Max}$ with $m_{\chi}$ and $k_{f_{\chi}}$. We find that all the above-mentioned parameters scale with the mass of the dark matter particle  $m_{\chi}$ in the same pattern for a given $k_{f_{\chi}}$ and follow an universal parametrization.
 We  also examine the dimensionless tidal deformabilty of $1.4M_{\odot}$. This investigation leads to narrowing  down the admissible range of dark matter fermi momentum and mass.

\bibliography{ref}  
\bibliographystyle{JHEP}

\end{document}